# Comparative analysis of criteria for filtering time series of word usage frequencies

Inna A. Belashova, Vladimir V. Bochkarev

Kazan Federal University
420008, Russia, Kazan, Kremlyovstaya 18
inkin91-91@mail.ru, vbochkarev@mail.ru

**Abstract.** This paper describes a method of nonlinear wavelet thresholding of time series. The Ramachandran–Ranganathan runs test is used to assess the quality of approximation. To minimize the objective function, it is proposed to use genetic algorithms - one of the stochastic optimization methods. The suggested method is tested both on the model series and on the word frequency series using the Google Books Ngram data. It is shown that method of filtering which uses the runs criterion shows significantly better results compared with the standard wavelet thresholding. The method can be used when quality of filtering is of primary importance but not the speed of calculations.

## 1 Introduction

Recently, a new tool for studying the dynamics of languages has become available. The Google Books Ngram corpus was created on the basis of a great number of digitized printed sources which have been published since the 16th century. It contains data on the frequency of words and phrases in different years for 8 languages [1].

Most of the printed materials were drawn from over 40 university libraries around the world. Each page was scanned and then digitized by means of optical character recognition (OCR). Publishers provided additional materials in physical and digital form and along with the libraries provided information describing the date and place of publication. Cultural trends can be quantitatively investigated via computational analysis of this corpus as frequency of word usage depends on various social factors [1]. The corpus is also used for creation of automatic translation systems and investigation of language evolution.

Frequencies of words, especially rare ones, fluctuate strongly. Therefore, it necessary to perform filtering of frequencies series to distinguish significant outbursts of the use of words associated with various social, political, and cultural processes, from random fluctuations.

The Google Books Ngram Viewer service created by the developers of the corpus uses moving average for this purposes. This, however, results in distortion of the series, and impedes the analysis of linguistic changes associated with various historical events. Using linear smoothing filters also shows unsatisfactory results.

Nowadays filtering methods based on statistical significance are wieldy used, in particular wavelet-thresholding. The method of wavelet-thresholding is based on the criterion of the minimum mean square error and is successfully used for the series which fluctuation probability distribution is close to the normal law. This approach can be incorrect in case of the series which distribution differs significantly from the normal one. The maximum-likelihood criterion is more universal than the methods mentioned above but requires knowing the law of time series value distribution. In practice, the distribution law is not always known a priori.

For example, it is often accepted that frequency distribution (at least the distribution of rare words) is governed by Poisson's law [3]. Filtering of time series with Poisson distribution was considered in number of works (see, for example, [4]). Nevertheless, there are a number of reasons to believe that a more complex model is required to describe the frequency distribution of words than the Poisson law. Thus, it is shown in [5] that the average value of the modulus of the relative change in frequency of words over an annual period is well approximated by a power function of the frequency of the words $f$. The following relationship is fulfilled in a wide range of frequencies for the English language corpus of the Google Books Ngram.

$$\left\langle \frac{|\Delta f|}{f} \right\rangle \sim f^{-0.316} \qquad (1)$$

Such behaviour of the frequency increments is most likely caused by the mutual influence of the authors on each other, and contradicts the assumption of the Poisson law for the frequency of words.

Thus, the development of filters that use robust quality criteria is an urgent task. In this paper, one of the possibilities of applying runs test is considered.

## 2  Method

The major challenge is to find a representation of the analysed data as a wavelet series with a small number (compared to the number of samples) of terms.

$$y(t) = \sum_{i \in I} c_i \psi_i(t) \qquad (2)$$

Here $y(t)$ is a filtered time series, $\psi_i(t)$ is a wavelet function corresponding to some scale and shift parameters (which are encoded by the multi-index $i$), $c_i$ is the corresponding coefficient, $I$ is a set of nonzero wavelet coefficients. It is assumed that the size of the set $I$ is much smaller than the length of the approximated time series.

As in the case of standard wavelet-thresholding, it is necessary to choose which terms of the wavelet series will be used for the approximation and estimate the optimum values of the corresponding wavelet coefficients. If the obtained approximation is good, the approximation errors are weakly correlated and don't form long runs (the sequences of points for which the error has the same sign), thereby allowing us to separate significant details of the times serious from the random ones.

The Ramachandran–Ranganathan runs criterion [6] is used in this paper. In accordance with this criterion, the objective function is written as the sum of the squares of the lengths of the runs $l_i$:

$$R = \sum_i l_i^2 \qquad (3)$$

We have used the Ramachandran–Ranganathan criterion because it is more powerful than other runs criteria, such as the Wald-Wolfowitz criterion. Thus, the calculation of the objective function for a given set of nonzero coefficients $I$ and values of the coefficients $c_i$ is performed in the following order:

- The approximation (2) is calculated and the approximation errors are found;
- Runs are identified and their length is determined for a number of errors.
- Statistics $R$ is calculated by the formula (3).

Since $R$ is a discrete function, determining the value of the coefficients $c_i$ by the criterion (3), only interval estimations can be obtained for them. If it is required to obtain concrete numerical values for $c_i$, we need to modify expression (3), replacing $R$ by a continuous function with similar properties. It should be note that to do it, the length of the runs $l_i$ can be written as follows:

$$l_i = \left| \sum_{t \in S_i} \text{sign}(\varepsilon_t) \right| \qquad (4)$$

Here, $S_i$ is the set of samples forming the $i$-th runs, and $\varepsilon_t$ is the approximation error at instant of time $t$. We can replace the function sign(x) in a given expression by a monotonically increasing smooth function that takes values from -1 to 1, for example, tanh(x). Thus, we obtain a «soft» runs criterion with statistics.

$$\tilde{R} = \sum_i \left[ \sum_{t \in S_i} \tanh \frac{\varepsilon_t}{\lambda} \right]^2 \qquad (5)$$

In this case, the value of the scale parameter $\lambda$ included in this expression can be chosen close to the expected level of additive noise.

To minimize the objective function $R$ it is expedient to use stochastic optimization algorithms. In this paper, a genetic algorithm is used to find the minimum. The positive features of genetic algorithms are their applicability to both continuous and discrete functions (which is especially important in our case), as well as the possibility of finding a global minimum error [7].

Construction of genetic algorithms begins with generation of populations containing a given number of chromosomes. The chromosomes evolve in the process of many iterations (generations). Similarly, the chromosome is evaluated during each iteration, in other words, the value of the objective function is calculated for the given chromosome. The next generation, called the descendant, is created with the help of two operators - the crossing operator (crossing-over) and the mutation operator. In the first case, the descendant is generated from two chromosomes by crossing the parents according to a given rule (one-point, two-point crossing-over), in the second case, the descendant is obtained by randomly changing the gene of one chromosome. After that, a new offspring is generated from the parental individuals and descendants selected according to the values of the objective function, and the remaining individuals are removed to maintain the population size constant.

The algorithm of filtering using genetic algorithms can be divided into two key blocks:

- Defining a new set of non-zero coefficients;
- Varying values of non-zero coefficients to minimize the statistics of the Ramachandran–Ranganathan runs criterion.

The set of non-zero coefficients of the wavelet decomposition is given by the chromosome. The chromosome is a string of zeros and ones, where «0» corresponds to the coefficients set to zero, and «1» corresponds to non-zero coefficients in formula (1). The genetic algorithm varies the positions of «1» in the chromosome, and there is a set at which the R statistic is minimal.

The most important condition for optimization is the constancy of the number of non-zero coefficients. It should be taken into account when developing the function of crossover and mutation, and also when creating the initial population. A uniform crossover was used in this work. The mutation function was developed taking into account the conservation of the number of non-zero coefficients: if «0» has changed to «1» in the chromosome, an inverse operation occurs in some other locus.

The so-called continuous genetic algorithm is used at the stage of varying the values of non-zero coefficients [7, 8]. Here the chromosome is no longer a bit string, but a vector of non-zero coefficients.

By setting the run of the value of non-zero coefficients, each of these coefficients is modified. Initial population is formed. In this case, crossing-over will occur in a completely different way than it was previously described. Here a mixed crossover is used, in other words, one descendent from the interval is generated by two progenitors:

$$[\min(x, y) - \alpha\Delta, \max(x, y) + \alpha\Delta], \quad \Delta = \max(x, y) - \min(x, y) \qquad (6)$$

Here, *x, y* are the values of the coefficient of two progenitors, and the parameter *α* is a small number that determines how much the coefficient of the descendent can go beyond the interval [*x, y*].

When specifying the mutation function, it is necessary to ensure that the coefficient is not zeroed. After that, calculation of the optimized function is performed. The operation is repeated until the algorithm stop condition is satisfied.

Often, a stop condition is used that stops the algorithm if the average relative change of the best value of the quality function is less than the specified limit accuracy after the specified number of iterations or equals to it. This condition was used in this paper.

## 3 Testing on model series

The proposed filter was tested on model series with a different law of distribution of variations. To perform testing, we used the example «Bumps» from the Wavelet Toolbox of the MATLAB package. The series with normal, Poisson and stable distributions (with the parameter α equal to 1.3), which mathematical expectation coincided with the selected test signal, were generated. Thus, the noise was additive for normal and stable distribution. For the case of the Poisson law, the numbers $n(t)$ were generated for each time $t$ in accordance with the distribution.

$$P(n,t) = \frac{\lambda^n(t)}{n!} e^{-\lambda(t)} \qquad (7)$$

The parameter of the Poisson law $\lambda(t)$ changed in proportion to the signal from the example «Bumps», but a constant was added to it so that all values were positive. In this case, the minimum and maximum values of $\lambda(t)$ were equal to 0.5 and 15, respectively. The three selected distributions are very different in their properties, which makes it possible to compare the filtering algorithms in different conditions.

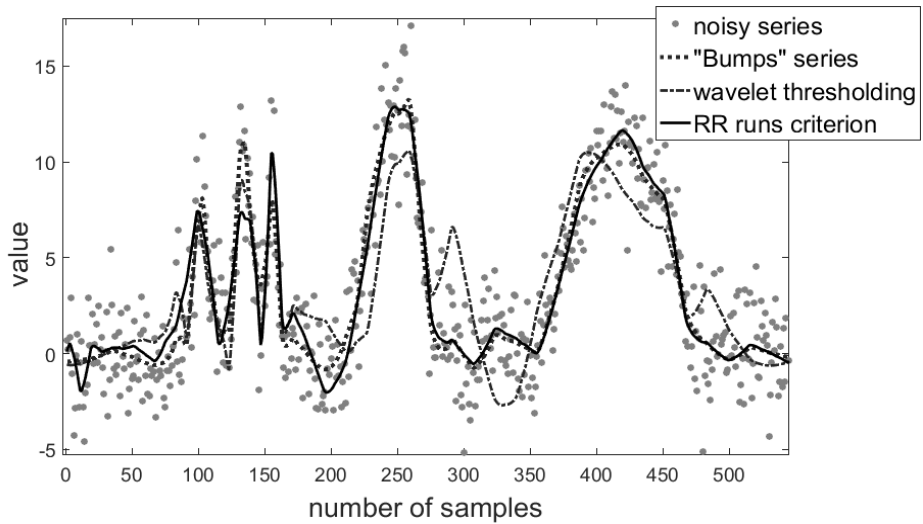

**Fig. 1.** Filtering of a noisy example «Bumps» (normal distribution). The dashed line shows the original «Bumps» series, the dots represent a noisy row, and the dash-dotted line shows the results of wavelet-thresholding, the solid line shows the results of filtering by the runs criterion

Wavelet-thresholding (by the minimax rule) and filtering by the described algorithm were performed for each generated series. Hereinafter, we used a wavelet basis based on the symlet of the third order, and the maximum possible number of decomposition levels. One of the cases is shown in Figure 1.

In this case, the thresholding algorithm leaves 35 non-zero coefficients out of a total of 1068 (that is, about 3.3%). When filtering using the runs criterion, we leave the same number of non-zero coefficients. It can be seen that the curve obtained by the runs criterion is much closer to the original signal than the curve obtained using wavelet thresholding (in particular, even the number of pulses in the signal is incorrectly determined for the latter curve).

Based on the results of processing of 50 random series (for each of the distributions), root-mean-square deviation of the filtered signal from the original signal «Bumps» was calculated. In this case, the standard deviation of the noise σ for the normal distribution was 2 (Since the standard deviation for the «Bumps» series is 3.77 and the signal-to-noise ratio is 3.56) and the scale parameter of the stable distribution was chosen equal to 1. The results for the three distributions are presented in the Table 1.

**Table 1.** Comparative RMS values of the filtered source signal for two quality criteria

|  | Normal distribution | Poisson distribution | Stable Distribution ($\alpha=1.3$) |
|---|---|---|---|
| Wavelet thtesholding | 0.6792 | 0.7846 | 1.8670 |
| Runs criterion | 0.3944 | 0.5064 | 0.8794 |

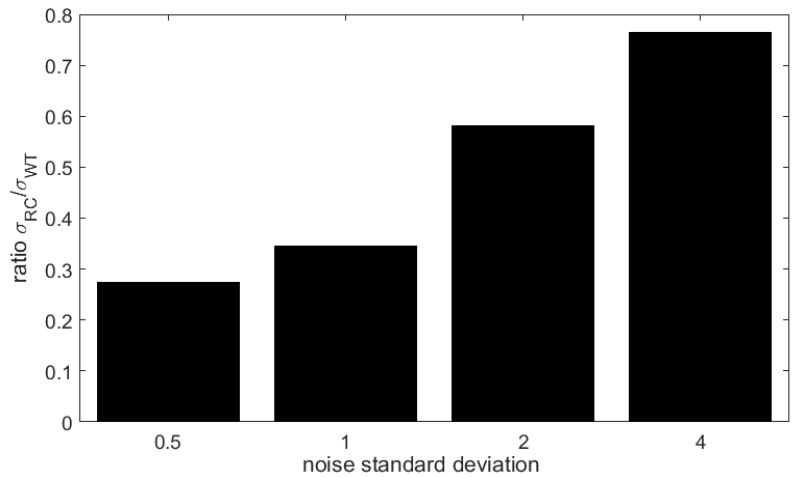

**Fig. 2.** Ratio of the root-mean-square error obtained after filtering by the runs criterion (RC) to the error obtained after using wavelet-thresholding (WT) for different noise levels

Thus, the results of statistical modelling demonstrate the advantage of filtering by the runs test over the standard wavelet-thresholding. It is particularly interesting that a significant improvement was obtained even for the case of a normal distribution of fluctuations.

The question arises: how the relative efficiency of the two filtration methods changes for the case of a normal distribution for different noise levels. Figure 2 shows how the ratio of filtering errors (i.e., the standard deviation of the filtered series from the original «Bumps» series) depends on the root-mean-square noise level for the compared methods. Error values obtained after filtering by the runs criterion were divided by the error values obtained after wavelet-thresholding, so the smaller values of the ratio mean a more significant advantage of the runs criterion over wavelet-thresholding. As can be seen from the figure, filtering by the runs criterion yields particularly good results at a moderate noise level. As for high noise series, the results obtained using the two methods come close to each other.

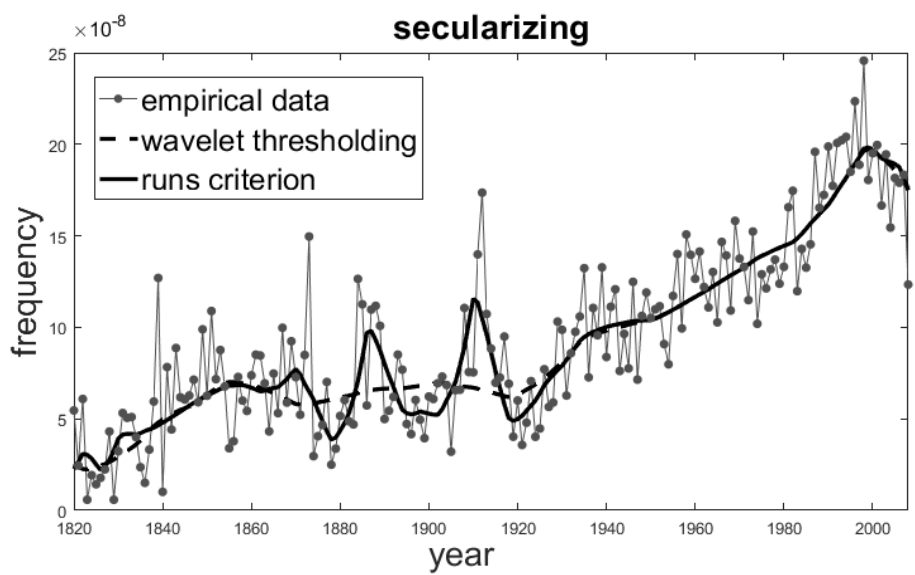

**Fig. 3.** A series of relative frequencies for the word «secularizing». The solid line shows the results of filtration by the runs criterion, the dashed line shows the series obtained using wavelet-thresholding

## 4     Filtering of series of words frequencies

Processing of series of words frequencies was also performed using the Google Books Ngram data (the 2009-year version of the corps was used). To carry out the analysis, words were randomly selected from the group of 100 thousand most frequently used words of the English language. We used series of average annual relative frequencies for the period 1800-2008. As in the case of model series, wavelet-thresholding (by the minimax rule) and filtering according to the runs test were performed (for each series). Figure 3 shows an example of filtering a number of frequencies for a word «secularizing». In this example, the approximating wavelet series contains 19 non-zero coefficients (of the total number 215).

It can be seen that application of the runs criterion allows us to obtain better approximation than wavelet-thresholding. The series obtained using the compared methods mainly diverge between 1880 and 1930. Significant frequency variations of the word «secularizing» are observed during this period. These variations disappear when using wavelet-thresholding, but are clearly seen when filtering by the runs criterion is used.

It should be emphasized once again that both of the compared approximations are wavelet series with the same number of non-zero coefficients.

Figure 4 shows an example of filtering a number of frequencies for the word «Hiibner» (this word can mean a surname, and besides it is used as the name of one of the species of butterflies) This example shows that wavelet-thresholding didn't cope with the task and the method based on the runs criterion reproduces the frequency trends well in the 1840-1970 interval. For example, let us pay attention to the sharp frequency jump in 1855, after which the frequency no longer decrease to the former small values. After conducting wavelet-thresholding, this jump is replaced by a smooth rise, which is a serious qualitative distortion of information. On the contrary, the suggested method correctly approximates the jump region.

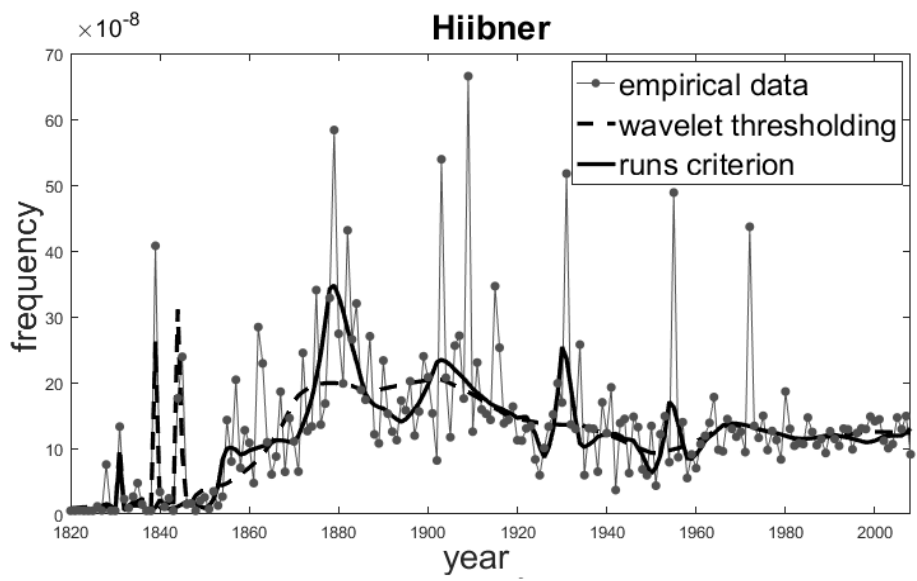

**Fig. 4.** A series of relative frequencies for the word «Hiibner» The solid line shows the results of filtration by the runs criterion, the dashed line shows the series obtained using wavelet-thresholding

There are some possible reasons why the approximation obtained by the wavelet-thresholding is so regrettable. Sharp peaks are observed in 1839 and 1844-1845 years. If these peaks are not taken into account, they will make a considerable contribution to the quadratic error. Therefore the thresholding algorithm considers them significant, «expending» several wavelet coefficients (out of total number of 19) on their reproduction. On the contrary, dropping such peaks does not contribute to statistics (3) and the method based on the runs criteria removes them quite reasonably. On the other hand, practical tasks are diverse and in some cases peaks can be important to the researcher. In this case, the runs criterion should be combined with a criterion that penalizes the drop of peaks. This can be realized by adding the corresponding term to the formula (3). Due to the universal nature of the genetic algorithm, it will not require sophistication of the calculation scheme.

A total of 70 examples were processed. As contrasted with the model series, in this case, we do not know the true values of the word frequencies in any given year. We only have the values obtained using the books included into the corpus, and therefore containing the error. In such situation, it is difficult to quantify how far one of the methods coped with the filtration task better than other methods. Therefore, the quality of the filtering was assessed visually for each of the examples. Difference in quality of filtering can be assessed as insignificant in 12 cases (17%). Filtering by the runs criterion gives better results than wavelet-thresholding in 31 cases (44%). And the advantage of filtering by the runs criterion can be assessed as very significant in 17 (39%) cases (the example shown in Figure 4 can be included into the latter group).

## 5    Conclusion

A method of nonlinear wavelet thresholding of time series is proposed in this paper. Signal representation in the form of a wavelet-series and the Ramachandran–Ranganathan runs criterion are used in this method to assess the quality of approximation. To optimize the objective function, the application of genetic algorithms is considered. A comparison of the proposed method with the standard wavelet-thresholding on model time series is performed. At that, the cases when the series are governed by the normal, stable, and Poisson distributions were considered. The de-scribed method was applied to filtering the time series of frequencies of the use of words and phrases (using the Google Books Ngram corpus data). It is shown that the use of a filter based on the runs criterion produces significantly better results than the use of both linear frequency filters and wavelet thresholding. It is clear that this algorithm has no advantage over wavelet-thresholding in the speed of data processing. Thus, this method can be used when quality of filtering is of primary importance but not the speed of calculations.

This work was supported by the Russian Foundation for Basic Research, Grant no. 15-06-07402. The research of the second author was supported by the Russian Government Program of Competitive Growth of Kazan Federal University.

# 6 References

1. Michel J., Shen Y., Aiden A., Veres A., Gray M., The Google Books Team, Pickett J., Hoiberg D., Clancy D., Norvig P., Orwang J., Pinker S., Nowak M., Aiden E. (2011) Quantitative Analysis of Culture Using Millions of Digitized Books. Science. 331 (6014), pp. 176-182.
2. S. Mallat A Wavelet Tour of Signal Processing, Third Edition: The Sparse Way 3rd Edition. 832 pages, Academic Press. 2008
3. R. H. Baayen Word Frequency Distributions, Kluwer Academic Publishers, 2001
4. Vladimir V Bochkarev and Inna A Belashova Modelling of nonlinear filtering Poisson time series 2016 J. Phys.: Conf. Ser. 738, 012082, doi:10.1088/1742-6596/738/1/012082
5. Bochkarev V, Solovyev V, Wichmann S. 2014 Universals versus historical contingencies in lexical evolution. J. R. Soc. Interface 11: 20140841. doi: 10.1098/rsif.2014.0841
6. David M. Himmelblau Process Analysis by Statistical Methods. John Wiley & Sons Inc, 1970.
7. Riccardo Poli, William B. Langdon, Nicholas Freitag McPhee A Field Guide to Genetic Programming Paperback – 2008, 252 pages, Lulu Enterprises, UK Ltd
8. F. Herrera, M. Lozano, J.L. Verdegay Tackling RealCoded Genetic Algorithms: Operators and Tools for Behavioural Analysis. Artificial Intelligence Review 12: 265–319, 1998. 265.